\documentclass{article}
\usepackage{format}
\usepackage{latexsym}
\usepackage[english]{babel}
\usepackage{bm}
\usepackage[T1]{fontenc}
\usepackage[utf8]{inputenc}
\usepackage{ae}
\usepackage{graphicx}
\usepackage{subcaption}
\usepackage[graphicx]{realboxes}
\usepackage{epsfig}
\usepackage{color}
\usepackage{pgf}
\usepackage{pstricks,pst-grad}
\usepackage{float}
\usepackage{sidecap}
\sidecaptionvpos{figure}{c}
\usepackage[normalem]{ulem}
\usepackage{latexsym}
\usepackage{amsfonts}
\usepackage{amssymb}
\usepackage{amsthm}
\usepackage{amscd}
\usepackage{amsmath}
\usepackage{mathrsfs}
\usepackage{breqn}
\usepackage[bbgreekl]{mathbbol}
\usepackage{mathtools}
\usepackage{color}
\usepackage{colortbl}
\usepackage{framed}
\usepackage{tcolorbox}
\usepackage{verbatim}
\usepackage{pdfpages}

\usepackage{microtype}
\usepackage[inline]{enumitem}
\usepackage{verbatim}
\usepackage{placeins}
\usepackage[full]{textcomp}
\usepackage{wasysym}
\usepackage{appendix}
\usepackage[nolist,nohyperlinks]{acronym}
\usepackage{diagbox}
\usepackage{booktabs}
\usepackage{xr}
\usepackage{amsthm}
\usepackage{qcircuit}
\usepackage{physics}

\usepackage{hyperref}
\hypersetup{
    colorlinks=true,
    citecolor=black,
    linkcolor=red,
    linkbordercolor=blue,
}
\usepackage{csquotes}
\usepackage[backend=biber,sorting=none]{biblatex}
\usepackage[nolist,nohyperlinks]{acronym}
\begin{acronym}[MPTPS]
    \acro{QVAE}{quantum variational autoencoder}
    \acro{QML}{quantum machine learning}
    \acro{QNN}{quantum neural network}
    \acro{VAE}{variational autoencoder}
    \acro{PCA}{principle component analysis}
\end{acronym}

\begin{document}

\renewcommand{\figureautorefname}{Fig.}
\renewcommand{\equationautorefname}{Eq.}
\renewcommand{\sectionautorefname}{Sec.}
\renewcommand{\subsectionautorefname}{subsec.}
\renewcommand{\tableautorefname}{Tab.}

\title{Learning Disentangled Representations with Quantum Variational Autoencoders}
\author{%
Gaoyuan Wang$^{1,2}$ \quad Jerry Tan$^{3}$ \quad Mark Gerstein$^{1,2,3,4,5*}$ \\ \\
$^1$ Program in Computational Biology and Bioinformatics,\quad \\
$^2$ Department of Molecular Biophysics and Biochemistry,\quad  \\
$^3$ Department of Computer Science,\quad  \\
$^4$ Department of Statistics \& Data Science,\quad  \\
$^5$ Department of Biomedical Informatics \& Data Science,\quad  \\
Yale University, New Haven, Connecticut 06520, USA \quad  \\
\texttt{*Corresponding author: pi@gersteinlab.org}\\
}
\maketitle

\begin{abstract}
Variational autoencoders are powerful representation learning models that map complex data into low-dimensional latent spaces, enabling the discovery of interpretable and disentangled factors. Such representations can facilitate the interpretation and controllable generation of data describing complex scientific systems. Understanding how these factors are organized and encoded in latent space is therefore important for developing reliable representation learning models.
Recently, quantum variational autoencoders (QVAEs) have been proposed as quantum representation models, demonstrating informative latent representations and improved latent-space occupancy through quantum regularization. However, it remains unclear whether and how QVAEs can learn disentangled and interpretable latent factors. A key challenge in investigating quantum latent factors is that a small number of qubits spans an exponentially large Hilbert space, making the notion of an individual quantum latent dimension nontrivial.
Here, we investigate what constitutes an individual quantum latent dimension and whether it can encode a distinct factor.
We develop theoretical insights into quantum latent dimensions and support them with empirical studies on representative synthetic problems, including MNIST variants. Across three datasets, we demonstrate that QVAEs can discover factorized and semantically interpretable latent representations, with individual qubits functioning as meaningful latent factors. These results establish a foundation for understanding quantum latent spaces and their potential for structured and interpretable representation learning.

\end{abstract}

\section{Introduction}

Machine learning and AI-powered methods are increasingly used to extract scientific insights from complex, high-dimensional data. A key challenge in this setting is to learn patterns that reflect meaningful structure in the data rather than merely reproducing their statistical features. Representation learning provides a framework for mapping high-dimensional data into lower-dimensional latent representations that capture underlying factors of variation, offering a path toward understanding the mechanisms underlying the scientific system of interest \cite{PhysRevLett.124.010508,10.1109/TPAMI.2013.50,wang2024disentangledrepresentationlearning}. 
This approach has found broad applications across scientific domains, including physical systems \cite{PhysRevX.10.031056,doi:10.1126/sciadv.abd5084,cwb8-y25k, qu2026representationlearningspatiotemporalphysical}, molecular and protein modeling \cite{10.1021/acscentsci.7b00572,Sevgen2025,Gujral2025}, and biomedical data analysis \cite{Lopez2018,Piran2024,Yun2024}.
For example, in molecular and protein modeling, learned latent representations have been shown to capture meaningful biological and structural variation, including associations with Gene Ontology terms, protein families, phylogeny, protein function, and molecular structural features. Representation learning can facilitate dimensionality reduction, visualization, prediction, the identification of meaningful patterns, and generative modeling. When distinct factors are encoded in separate latent variables, the resulting representations can make these factors easier to interpret and independently manipulate. Understanding how such factors are organized and encoded in latent space is therefore important for developing reliable and scientifically meaningful machine learning models.

Variational autoencoders (VAEs)\cite{kingma2022autoencodingvariationalbayes,10.1561/2200000056} are a central class of representation learning models that learn low-dimensional latent representations while providing a framework for generative modeling. Their architecture consists of an encoder that maps inputs into a latent space and a decoder that reconstructs the inputs from the latent variables. In addition to the reconstruction loss, which encourages the reconstructed state to remain close to the input, VAEs impose a regularization penalty on the latent space. In classical VAEs, this regularization encourages the learned latent distribution to match a standard isotropic Gaussian prior, promoting a continuous and structured latent space with broad coverage. These properties can improve generalization and facilitate interpolation and generative modeling.
The $\beta$-VAE model \cite{higgins2017betavae} further strengthens this regularization and has been widely studied as a framework for learning disentangled representations. By increasing the relative weight of the regularization term, $\beta$-VAEs can encourage individual latent dimensions to capture distinct factors of variation while remaining relatively insensitive to other variations. Such disentanglement improves the interpretability and controllability of the learned representation and provides a means of connecting individual latent variables to distinct properties of the underlying data \cite{higgins2017betavae,burgess2018betavae,tschannen2018recentadvancesautoencoderbasedrepresentation,higgins2018definitiondisentangledrepresentations,pmlr-v97-mathieu19a,LIU2022102516,10579040,9947342}.

In Ref.~\cite{PhysRevA.111.042416}, a quantum variational autoencoder (QVAE) framework, $\zeta$-QVAE, was proposed. This model employs a quantum regularization term based on the maximally mixed state as the prior. Using a quantum encoder, QVAEs map quantum states from a high-dimensional Hilbert space to a lower-dimensional latent Hilbert space and reconstruct the original quantum state through a quantum decoder. At an optimal regularization strength controlled by the parameter $\beta$, $\zeta$-QVAE has been shown to achieve maximal utilization of the latent Hilbert space, better preservation of the input data structure, and improved performance on downstream tasks compared with unregularized quantum autoencoders. These results demonstrate that quantum regularization can strongly influence both the structure and utilization of the quantum latent space.

However, an important question remains unexplored: can QVAEs learn disentangled and interpretable latent factors, and if so, how are these factors encoded in the quantum latent space?
Addressing these questions is nontrivial. First, the notion of a latent dimension in a QVAE is far from straightforward and is fundamentally different from that in a classical VAE. In classical VAEs, each latent variable corresponds to an explicit degree of freedom and can, in principle, encode an individual generative factor. In contrast, a quantum latent state of $n$ qubits resides in a $2^n$-dimensional Hilbert space, while its physical representation and operations are defined at the level of $n$ qubits. Although the quantum latent state can be expressed in the computational basis, the exponentially large Hilbert space does not itself provide a natural correspondence between basis states, qubits, and meaningful latent factors. Consequently, we first have to address what should be regarded as an individual quantum latent dimension. Second, associating a learned latent variable with a meaningful scientific factor generally requires post hoc analysis based on correlations with known properties or perturbations in the data. This is because representation learning models, including classical ones, are typically data-driven and do not impose a first-principles correspondence between latent variables and the underlying factors of a scientific system.

These challenges make a general closed-form characterization of the learned latent space difficult. We therefore combine theoretical analysis with controlled empirical experiments designed to expose the relationship between quantum latent degrees of freedom and known factors in the data. In particular, we study representative synthetic problems and MNIST \cite{726791} variants with known underlying factors. We show that individual qubits can serve as meaningful latent dimensions and that $\zeta$-QVAE can discover factorized and semantically interpretable latent representations in an unsupervised setting. These results provide insight into the structure and behavior of quantum latent spaces and establish a basis for interpreting quantum latent representations in quantum machine learning and, more broadly, scientific data analysis.

\section{Quantum Representation Learning}
In this section, we provide our intuitions and insights towards
quantum representation learning with QVAEs. In \autoref{sec:se2QAE} we describe our intuition of what happens in the quantum autoencoder without regularization. Then in \autoref{sec:sec2regu} and \autoref{sec:sec2dis}, we present our insight into the impact of regularization and why the latent information bottleneck strengthened by the regularization leads to disentangled factors aligned with computational qubits.

\subsection{Quantum Autoencoders Without Regularization}
\label{sec:se2QAE}
The input dataset for QVAE is a set of quantum states $\{\ket{\psi_u}\}$. These quantum states can either be of quantum origin or computed from a classical origin by doing a standard quantum state encoding, such as amplitude encoding, on the classical input $x_u$ via $\ket{\psi_u}=\textit{Amplitude}(x_u)$. These $d$-qubits quantum states live in a $2^d$-dimensional Hilbert space, which can be used to represent a $2^d$-dimensional classical vector.
QVAE in the absence of regularization, i.e., at $\beta=0$, is effectively a Quantum Autoencoder (QAE) \cite{Romero_2017}. Here, we discuss what happens when a QAE is trained to optimize the reconstruction loss alone.

At the latent bottleneck of a QAE, the $d$-qubit quantum space can be seen to be divided into two subspaces: $k$ latent qubits in the Hilbert space $\mathcal{H}_A$ that is to be preserved and $m = d - k$ trash qubits in the Hilbert space $\mathcal{H}_B$ that is to be discarded. 
Since QAE's architecture utilize partial trace over trash qubits for discard and zero padding to extend dimensionality back to the original one for reconstruction, the preserved subspace is spanned by the dimensions with the form $\ket{...}_A \otimes \ket{0}^{\otimes m}_B$
while the discarded subspace is spanned by those of the form $\ket{...}_A \otimes \ket{j \neq 0}^{\otimes m}_B$. 
The optimization of parameters in QAE is to find the best strategy on what to put into the preserved and discarded subspace. 

There are several mechanisms on what QAE can bring to the trash subspace.
The first one is to remove redundancy across data points by storing common features globally rather than repeatedly encoding them for each individual sample. The encoder can rotate low-variance dimensions across the dataset into the trash subspace. Since these dimensions contain nearly constant information shared across the data ensemble, they do not require instance-specific encoding and can instead be captured by the encoder and decoder parameters. During reconstruction, these globally stored features can be recovered by the decoder from the padded zero states in the trash subspace.
If the variance in these dimensions is truly small, this provides very small information loss, regardless of the magnitude of these dimensions.

The second one is when two or more dimensions are highly correlated. In this case, QAE can keep one of these dimensions in the preserved subspace and move the rest to the trash subspace. This mechanism can also result in minimal information loss.

The third one is to discard the insignificant eigenbases by rotating then into the trash subspace. This would generate reconstruction error, as instance-specific information in these dimensions are lost. But only sacrificing low significance dimensions helps reducing reconstruction error.
To understand this, we construct the ensemble density matrix:
\begin{equation}
\rho_{\text{ens}} = \sum_u p_u \ket{\psi_u}\bra{\psi_u},
\end{equation}
where each $\ket{\psi_u}$ represents one of the $N_{\text{ens}}$ quantum input data points, and $p_u$ their probability in the ensemble. 
Diagonalizing $\rho_{\text{ens}}$ and formulating it in its eigenbasis $\ket{e_j}$, we have
$\rho_{\text{ens}} = \sum_{j=1}^{2^d} \lambda_j \ket{e_j}\bra{e_j},$
where $\lambda_j$ are eigenvalues in descending order.
Each individual data point $\ket{\psi_u}$ can be decomposed in the same eigenbasis:
\begin{eqnarray}
    \ket{\psi_u}\bra{\psi_u}=\sum_{j=1}^{2^d}c_{uj}\ket{e_j}\bra{e_j}\quad
  \text{and} \quad\sum_{u=1}^{N}c_{uj}= \lambda_j.
\end{eqnarray}
Eigenvectors $\ket{e_j}$ with small eigenvalues $\lambda_j$ are characterized by small average overlap with the states in this ensemble, meaning they contribute small to most individual data points. Conversely, eigenvectors with large eigenvalues are those in which the states in this ensemble on average have significant amplitude.
The typical subspace spanned by eigenvectors with large eigenvalues are frequently occupied by states in the ensemble. Atypical subspace, spanned by eigenvectors with small eigenvalues are less occupied by ensemble states. 

To summarize the mechanism of QAE, optimizing on the reconstruction loss incentivizes the encoder $U$ to approximate a transformation that stores ensemble-level information into the encoder and decoder, preserves per data point latent state carries the preserved instance-specific information are preserved, while discarding redundant information and insignificant information get discarded.

It is worth noting the connection between QAE and the Schumacher compression \cite{PhysRevA.51.2738,PhysRevA.54.2636}, a compression scheme rooted in quantum information theory, where the von Neumann entropy $S(\rho_{\text{ens}})$ sets the lower bound on the number of qubits needed to faithfully store long sequences drawn from the ensemble $\rho_{\text{ens}}$. The split between typical and atypical subspaces in the QAE echoes this logic, but the two schemes are not equivalent. Schumacher compression operates on the entire block of a sequence drawn from $\rho_{\text{ens}}$, allowing for joint encoding across the individual states in that sequence, whereas the QAE applies the same unitary to each instance independently. Thus, $S(\rho_{\text{ens}})$ provides a loose lower bound on the compression rate achievable by the QAE, with equality attained only in the degenerate case where the Schumacher encoding itself has no cross-instance entries and applies the same block transformation to each instance.

It is also worth noting the connection between QAE and \ac{PCA}. While the goal of QAE in compressing quantum data parallels that of classical autoencoders and \ac{PCA} on classical data, the underlying mechanism differs. \ac{PCA} is restricted to linear, orthogonal projections and equivalent to a linear classical autoencoder, while QAE can capture nonlinear relationships between dimensions through quantum phenomena such as entanglement. More fundamentally, \ac{PCA} diagonalizes the covariance matrix, so its eigenvalues directly measure variance, and low-variance directions are discarded. The QAE instead diagonalizes the ensemble density matrix, whose eigenvalues measure occupation probability, i.e., how strongly the ensemble's states overlap with a given eigenspace on average. In QAE, top-$k$ eigenvalue truncation (the third mechanism described above) alone does not guarantee the most informative latent states for reconstruction fidelity, since discarding only the atypical subspace does not handle redundant dimensions with large eigenvalues. Because QAE is trained end-to-end rather than obtained via closed-form eigendecomposition, it leaves room for regularization to guide the encoder toward the correct latent structure, which we explore in the following.

\subsection{Impact of Regularization}
\label{sec:sec2regu}
In representation learning, unlike QAE, the objective is typically not just to minimize reconstruction loss as in pure compression, but to learn meaningful and useful features from the data. Faithfully preserving noise is often of little value for downstream tasks; instead, it is usually preferable to discard noise and retain the underlying structure and informative factors.

In both classical VAE and QVAE, regularization is employed to force the models to use latent dimensions efficiently and independently.
In classical VAE, as proposed in Ref.\,\cite{higgins2017betavae}, applying a strong regularization has been shown to enable unsupervised learning of disentangled representations by strengthening the pressure for the approximate posterior over the latent variables to match the factorized latent prior.
In \cite{burgess2018betavae}, authors provided additional intuitions and theoretical justification for $\beta$-VAE's success in disentangled factor learning. They argued that the reduced the encoding capacity of the latent bottleneck encourages the learning of disentangled representations. They also explained why emergent latent variables align with interpretable generative factors: under a strong information bottleneck, the model has limited capacity and therefore must prioritize encoding only the factors of variation that most reduce reconstruction error. As a result, $\beta$-VAE tends to dedicate separate latent dimensions to distinct influential factors whose effects on the data differ, such as position, scale, or orientation. 

In QVAE, the prior, the maximally mixed state, penalizes correlations between preserved dimensions and encourages the preserved dimensions to be decoupled, thus achieving similar disentanglement affect on regularizing quantum latent space as regularization in $\beta$-VAE.
When regularization is applied to the objective, the model must balance reconstruction fidelity with additional constraints on $\mathcal{H}_A$, rather than optimizing reconstruction loss alone. This trade-off reduces reconstruction quality compared to the unregularized case for the same latent dimensionality $m$. But this worse reconstruction is consciously accepted to achieve disentanglement.

\subsection{Disentangled Factors in Quantum Representations}
\label{sec:sec2dis}
In classical VAEs, the latent space is usually a Euclidean vector space, and latent dimensions correspond naturally to coordinates within that space. 
In quantum systems, because the quantum latent space is a Hilbert space, the notion of latent dimensions is less straightforward. The key complication is that the dimension of the Hilbert space is exponential in the number of qubits. The maximally mixed state prior decouples all dimensions in the Hilbert space without seeing the boundary formed by computational qubits. But the fact, that parametrized gates in the circuits are operating on computational qubits introduces a structural preference for the emergent disentangled factors to align with qubits. 
When a factor is isolated on a single qubit, the gradient feedback is focused, involving less gates and parameters, providing a clear and efficient optimization path. In contrast, when a factor is distributed across multiple qubits, the feedback is scattered across many gates and parameters and entangling gates, creating a noisy training signal. Thus, learned disentangled factors tend to localize to individual qubits.

Concretely in this work, we decompose the latent quantum state into subsystems. Each subsystem is obtained by taking a partial trace over all qubits except the qubits of interest. These reduced states each represents a candidate latent dimension that may encode a disentangled factor. 
When choosing the size of the latent space—or equivalently, the number of latent qubits in a QVAE—we recommend starting from an estimate of the number of expected disentangled generative factors in the data. From there, the latent dimensionality can be adjusted empirically: too few dimensions typically result in poor reconstructions, while too many can introduce uninterpretable and badly factorized latent dimensions. In practice, this selection can be refined more systematically by identifying the threshold between informative and noisy latent dimensions.

Similar to the works in classical VAE \cite{higgins2017betavae}, we use an information-based metric to quantify the correspondence between qubits and disentangled factors, i.e., how much information about a target factor (e.g., rotation angle) is stored in a particular latent subsystem. Specifically, we train simple classical or quantum regressors using only the reduced states as input. For the quantum regressor, the input consists of the density matrix of the candidate reduced states, and the model is trained to predict the corresponding generative factor from this input. For the classical regressor, we first map the reduced states into its coordinates $(x,y,z)$ on the Bloch sphere by computing expectation values of the Pauli operators, and then predict the target factor from these coordinates.

In both cases, we measure the information content via prediction performance (accuracy for multi-class classifiers and mean-squared error (MSE) for continuous regressors) of the target factor. Lower error or higher accuracy implies higher information content. To calibrate the scoring range for MSE, we introduce a baseline representing no information: we generate completely random quantum states and evaluate the regressors on them. All reported losses are normalized by this random-state baseline.

\section{Learning Disentanglement Factors in MNIST Images}
We first investigate the quantum representations learned by $\zeta$-QVAE on images from the MNIST handwritten digit dataset. To identify disentangled factors in the quantum representations, we obtain single-qubit subsystems of the quantum latent state by taking the partial trace over all other qubits.
The original images have a resolution of $28 \times 28$ pixels, which we downsample to $8 \times 8$ pixels and encode them onto 6 qubits using amplitude embedding. We conduct two complementary studies, each consisting of 1000 images balanced across the relevant classes. The first study in \autoref{sec:raw01234_result} focuses on the digits 0, 1, 2, 3, and 4, providing a discrete-class scenario. The second study in \autoref{sec:rotation_result} examines digits 3 and 7 with randomly applied rotations, introducing a continuous, generative factor. 

When illustrating latent representation distributions, the coordinates $(x, y, z)$ of certain qubit for a given latent state are given by the expectation values of the corresponding Pauli operators $(P_x, P_y, P_z)$ acting on that qubit.
Because the input consists only of real-valued pixel intensities and the chosen ansatz (Real Amplitude) does not introduce complex amplitudes, the expectation value of $P_y$ is always zero. As a result, the distribution of latent states is confined to the $P_x$–$P_z$ plane. We sort the indices of the latent qubits according to their association with the target disentangled factor, such that qubit 0 denotes the qubit most strongly associated with this factor.

Results at $\beta=0$ provide the baseline with no regularization, corresponding to a simple quantum autoencoder.
Statistical metrics (e.g., accuracy and loss) were computed by averaging each $\beta$ and qubit over nine independent runs. 
Additional details on dataset preparation can be found in Method section.

\subsection{Results on Raw MNIST Digit Images}
\label{sec:raw01234_result}
\begin{figure}[htbp]
    \centering
    \includegraphics[width=1\linewidth]{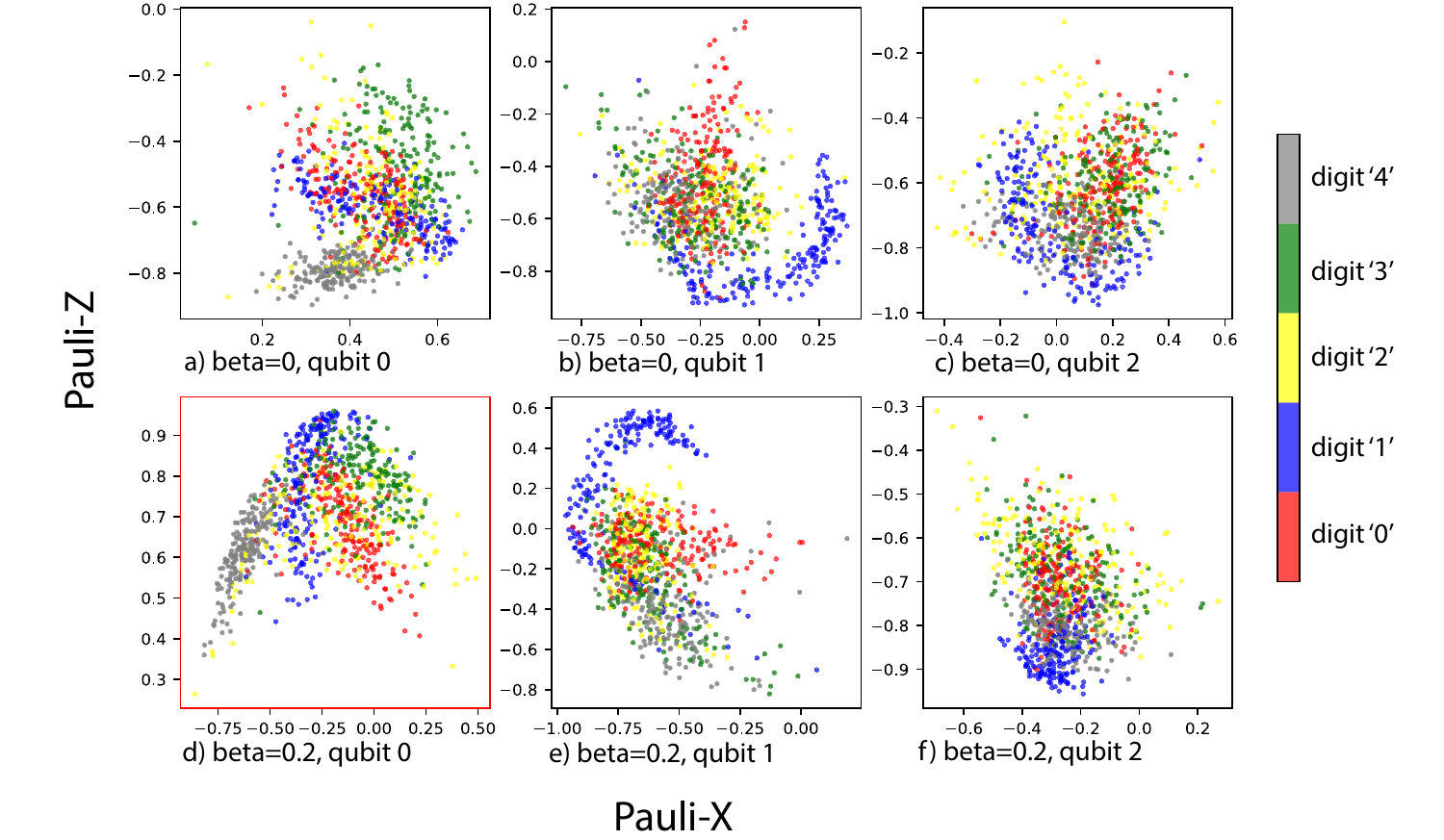}
    \caption{Latent representation distributions. At $\beta = 0.2$, $\zeta$-QVAE encodes the majority of class information in qubit 0, demonstrating successful disentanglement of latent factors.}
    \label{fig:latent_space_raw01234}
\end{figure}

We apply $\zeta$-QVAE to $8 \times 8$ images of digits 0, 1, 2, 3, and 4 using three latent qubits.
To examine how latent states from different classes are distributed in the latent Hilbert space, we presented a latent scatter plot, where we plot the location of each latent state for each latent qubit in \autoref{fig:latent_space_raw01234}. The color of each point denotes the digit class of the corresponding input image. 

For qubit 0, which carries the majority of the class information, setting $\beta = 0.2$ (\autoref{fig:latent_space_raw01234} panel d) yields a well structured and continuous latent representation. In this regime, the digit classes transition smoothly into one another, forming a continuous distribution with limited overlap between classes. 
This behavior is consistent with the latent space structure observed in classical VAEs trained on MNIST, as commonly illustrated through either latent scatter plots or decoder-based manifold visualizations. Such structure is reported in standard implementations (e.g., the Keras VAE tutorial \cite{keras_vae_tutorial}), in the original variational autoencoder formulation \cite{Kingma2013AutoEncodingVB}, and in subsequent works where latent distributions are analyzed for comparison with more advanced models (e.g., Fig.\,2(c) of Ref.\,\cite{Makhzani2015AdversarialA}).

In contrast, at $\beta = 0$ without regularization (panel a), our latent distribution of qubit 0 is significantly less structured and exhibits greater overlap between classes.
Panels b), c), e), and f) of \autoref{fig:latent_space_raw01234} show the remaining latent qubits. These qubits display little correlation between their coordinates and the digit classes, indicating that they carry minimal class-related information. This demonstrates that $\zeta$-QVAE primarily utilizes qubit 0 to encode class information, thereby achieving a clear disentanglement of latent factors.

To quantify the degree of disentanglement and assess how well the latent space is structured, we train a simple classical neural network that takes as input the coordinates of a latent qubit and predicts the digit class. If a qubit does not encode class-related information, the classification accuracy of this network should be around 0.5 comparable to random guessing. Conversely, higher accuracy indicates that the qubit carries more class information.

\begin{figure}[htbp]
    \centering
    \includegraphics[width=0.5\linewidth]{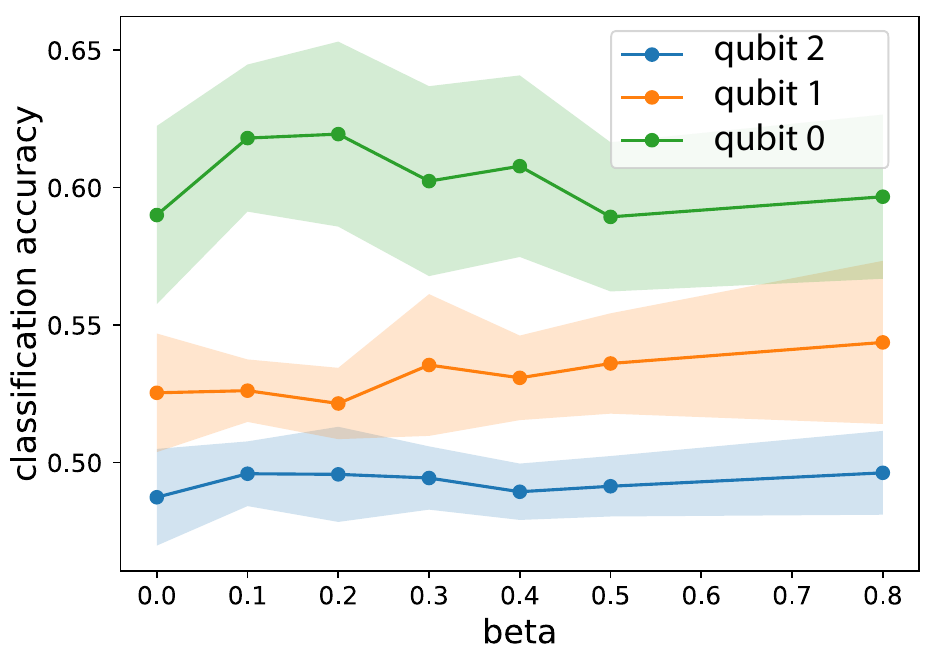}
    \caption{The multi-class classification accuracy of a simple predictive neural network quantifies the amount of class information encoded in each qubit. Qubits 1 and 2 perform at near–random-guess levels, while qubit 0 encodes the majority of the class information. Shaded areas indicate the error range.}
    \label{fig:accu_raw01234}
\end{figure}

In \autoref{fig:accu_raw01234}, we report the multi-class classification accuracy as a function of $\beta$ for latent qubits 0, 1, and 2. Qubit 0 consistently achieves an accuracy well above 0.5, while qubits 1 and 2 remain close to 0.5. This confirms that the majority of class-related information is encoded in qubit 0, with the remaining qubits largely uncorrelated with the class label, providing quantitative evidence of successful disentanglement of the class-related latent factor.
Moreover, the classification accuracy for qubit 0 reaches its maximum at $\beta = 0.1$ and $0.2$, exceeding the performance observed at $\beta = 0$ (the case without regularization), while the accuracy of the remaining qubits remains independent of $\beta$. This indicates that the regularization enhances the structure of the target latent factor only along its associated dimension, further confirming the disentanglement of the latent factors.

\subsection{Results on MNIST Digits with Random Rotations}
\label{sec:rotation_result}
\begin{figure}[htbp]
    \centering
    \includegraphics[width=1\linewidth]{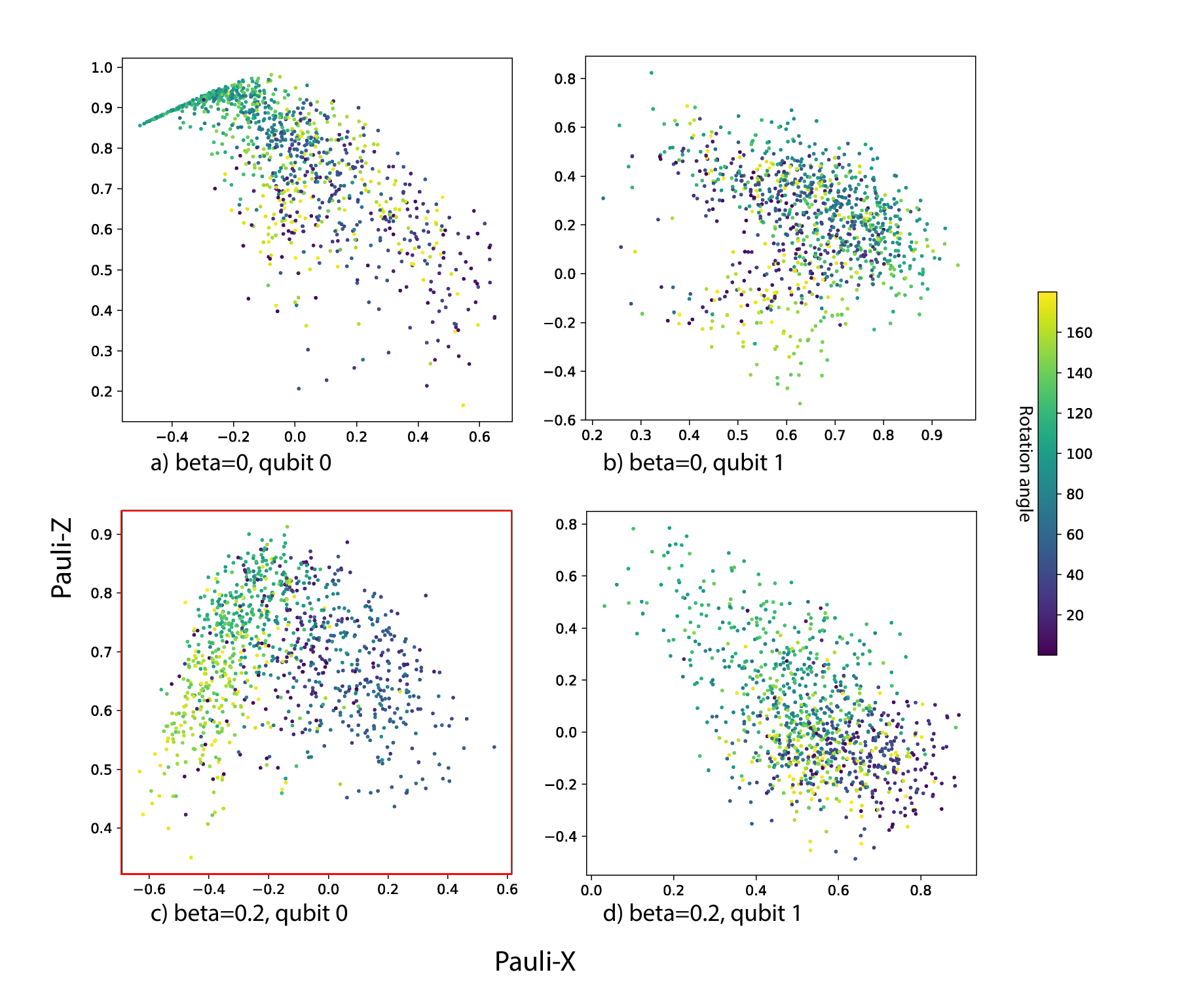}
    \caption{Latent distributions for rotated digits. At $\beta = 0.2$, qubit 0 aligns with rotation angles. Qubit 1 shows no correlation, indicating disentanglement of latent factors.}
    \label{fig:rotation_latent_space}
\end{figure}

We introduce a second artificial factor into the dataset by applying random rotations to the digit in each image.
We apply $\zeta$-QVAE to $8 \times 8$ images of randomly rotated digits 3 and 7 using two latent qubits. The locations of the latent states for each qubit are plotted in \autoref{fig:rotation_latent_space}, with a continuous colormap representing the rotation angle. The qubits are indexed such that qubit 0 is most strongly associated with rotation.
In panel c), qubit 0 at $\beta = 0.2$ exhibits a well-structured, smooth transition of latent state locations with respect to rotation angle. The nonlinear relationship between rotation angle and latent state positions highlights the QVAE’s ability to capture complex latent factors. At $\beta = 0$, qubit 0 only partially separates rotation angles around 100° (green) from the rest, but small (0°) and large (180°) rotations overlap. In contrast, qubit 1 shows little alignment with rotation in either regime, carrying minimal rotation-related information, which indicates successful disentanglement of the latent factors.

\begin{figure}[htbp]
    \centering
    \begin{subfigure}{0.48\textwidth}
    \includegraphics[width=1\linewidth]{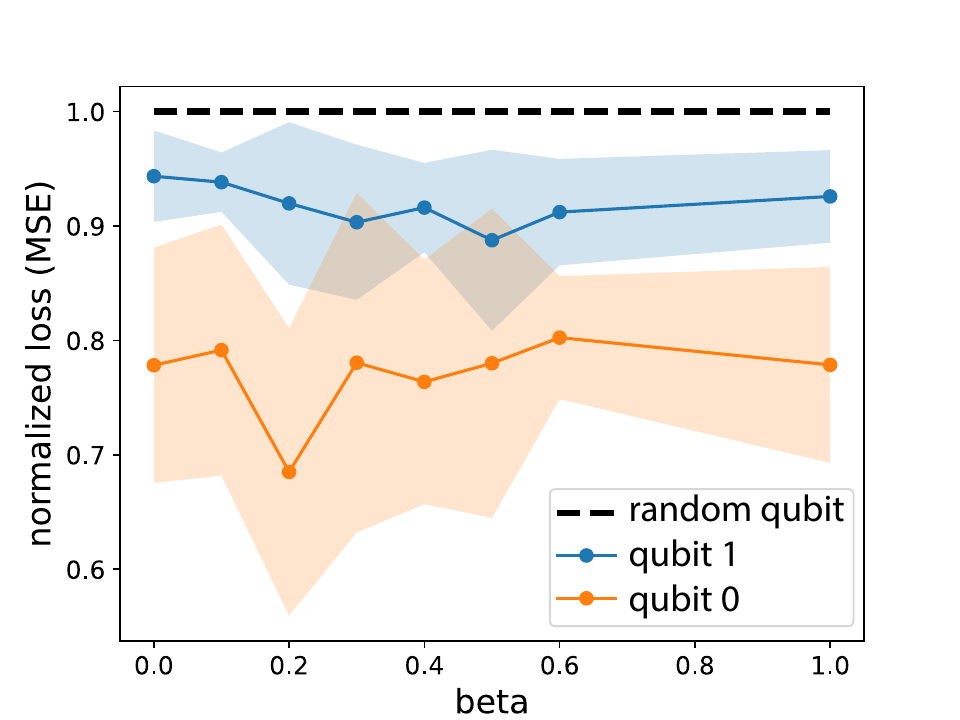}
    \caption{Classical Neural Network}
    \end{subfigure}
    \begin{subfigure}{0.48\textwidth}
    \includegraphics[width=1\linewidth]{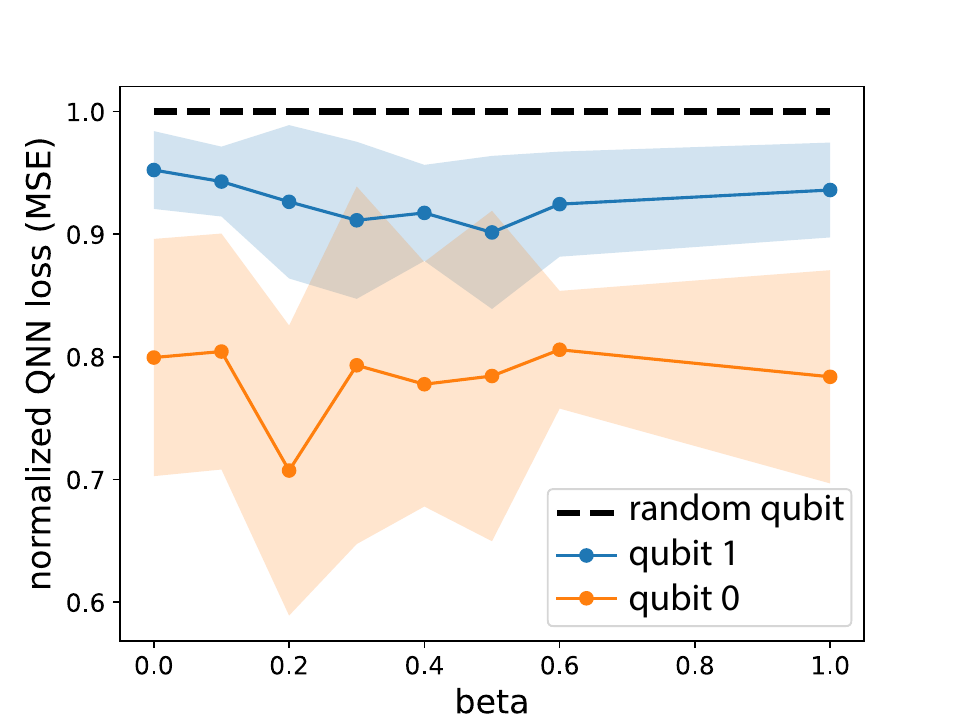}
    \caption{Quantum Neural Network}
    \end{subfigure}
    \caption{The MSE of simple predictive neural networks quantify the amount of rotation-angle information encoded in each qubit. Qubit 0 encodes the majority of the information, while qubit 1 performs near random. Shaded areas indicate the error range.}
    \label{fig:MNIST_rot_2dim}
\end{figure}

To quantify the information about rotation stored in each qubit, we train simple neural networks to predict the rotation angle from each latent qubit. For the classical network in \autoref{fig:MNIST_rot_2dim} panel (a), the $P_x$ and $P_z$ expectation values of the latent qubit states are used as input. For the quantum neural network in panel (b), the quantum state of the corresponding qubit is used directly.
\autoref{fig:MNIST_rot_2dim} shows the normalized mean squared error (MSE), scaled against a random qubit baseline (representing zero information), for both qubits. Qubit 1 exhibits MSE close to the baseline, while qubit 0 achieves a substantially lower MSE, indicating it stores significantly more rotation-angle information. We observe a clear dip in MSE for qubit 0 at $\beta = 0.2$, whereas qubit 1 remains unchanged. This demonstrates that the optimal regularization enhances the representation of the target factor along its associated latent dimension without affecting disentanglement of the remaining qubits.

\section{Learning Disentanglement Factors in Spectral Lines Dataset}
We continue investigating the quantum latent representations learned by $\zeta$-QVAE on the synthetic spectral lines dataset. The model receives as input 64-dimensional Gaussian spectral lines controlled by two continuous generative factors: peak position ($\mu$) and peak width ($\sigma$). Our goal is to assess whether the model learns these factors in an unsupervised manner. We do so by identifying disentangled quantum latent dimensions that align with the ground-truth factors. Additional details on the spectral dataset can be found in Method section.

To extract latent subsystems, we perform partial traces over all qubits except those of interest. Initial experiments with two latent qubits revealed two key findings: (1) the factor width is substantially more difficult to encode than the factor position, and (2) the latent bottleneck of two qubits was insufficient for accurate width encoding. The difficulty in encoding width arises because this factor affects all spectral dimensions non-uniformly, unlike position, which acts as a global shift. To address this, we increased to three latent qubits. When interpreting the qubit subsystems, we assume that one of the two factors will be encoded by one qubit, and the other factor will be encoded by the remaining two qubits. We systematically test all grouping of qubits and factors and found that for the majority of the $\beta$ values and across multiple independent runs with varying initialization parameters, the optimal grouping consistently allocated a single qubit to encode position and two qubits to encode width, reflecting the different in difficulty of the two factors.

\begin{figure}[htbp]
    \centering
\includegraphics[width=1\linewidth]{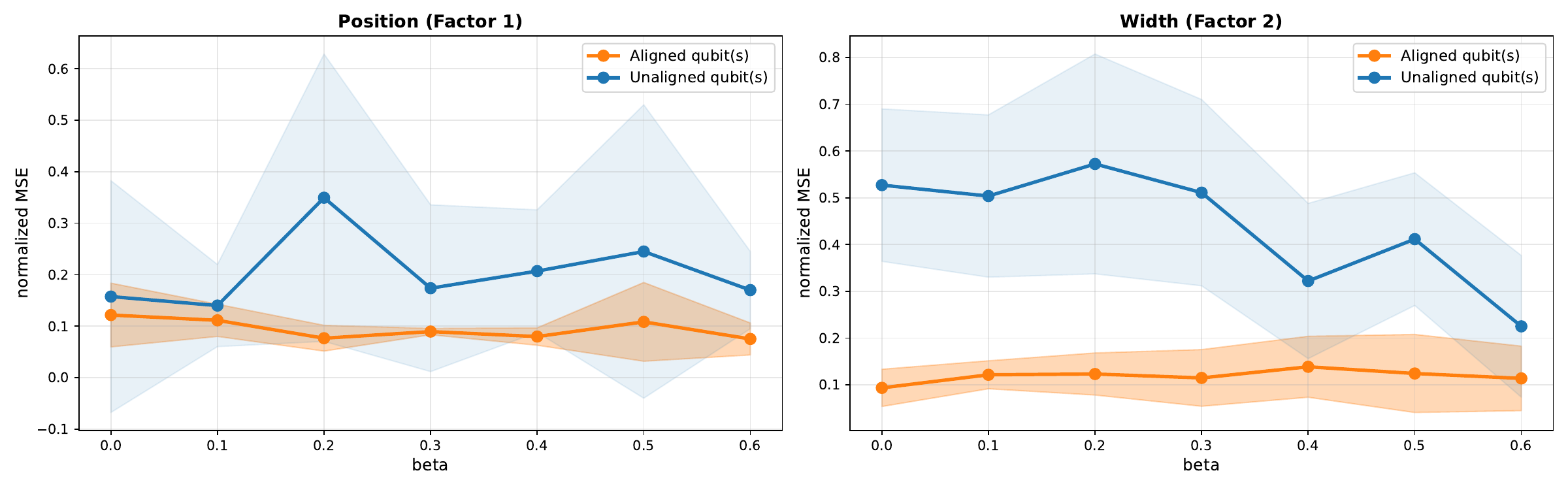}
    \caption{MSE of a simple classical neural-network predictor used to quantify the amount of generative-factor information encoded in each subsystem. Qubit 0 predominantly encodes position information, whereas qubits 1 and 2 primarily encode width information. Shaded regions indicate the error range across runs.}
    \label{fig:Spectral_mse}
\end{figure}

In subsequent illustration, for clarity, we reorder the qubit indices such that qubit 0 encodes position (factor 1) and qubits 1–2 jointly encode width (factor 2).
To facilitate comparison across settings, we normalize the MSE by a baseline MSE defined as the error obtained when the same regressor is trained using the corresponding number of random qubits and factor value. Accordingly, we consider four distinct baseline MSEs: one random qubit with factor 0, two random qubits with factor 0, one random qubit with factor 1, and two random qubits with factor 1.
We report the normalized MSE values averaged over nine independent runs for each $\beta$ value in \autoref{fig:Spectral_mse}. Here, the orange curve (aligned qubit(s)) uses as input the qubit identified as the one encoding the corresponding factor. The blue curve (unaligned qubit(s)) represents the MSE learned by the regressor using the qubits that were found to encode the other factor and thus not align with this factor.
In the left panel, we show the normalized MSE of the regressor predictions for the factor 1 (position). In most cases, the aligned qubit corresponds to a single-qubit subsystem, whereas the unaligned qubits generally correspond to two-qubit subsystems.
In the right panel, we present the normalized MSE for the factor 2, namely the width. In this case, the aligned curves predominantly correspond to two-qubit subsystems, while the single-qubit subsystems associated with this factor are shown as unaligned.
These results demonstrate clear disentanglement in factor learning. The position factor is consistently encoded more effectively by qubit 0, while showing substantially weaker association with the remaining subsystem qubits. Conversely, the width factor is primarily encoded by qubits 1 and 2, and can be inferred much less accurately from qubit 0.

For the factor position, the average MSE improves from 0.1217 at $\beta=0$ to 0.0766 at $\beta=0.2$, corresponding to an approximate $37\%$ reduction in error. At the same value of $\beta$, we also observe a significant decrease in the association between this factor and the remaining latent subsystem, indicating strong disentanglement at $\beta=0.2$.
On the other hand, for the width factor, the MSE of the aligned qubits does not improve significantly at nonzero values of $\beta$. Nevertheless, there remains a clear separation across all $\beta$ values between the association of this factor with the aligned qubits and with the unaligned qubit. Moreover, predictive accuracy of the aligned qubits, quantifying the amount of information about this generative factor stored in the qubit, fluctuates around a consistently informative level for all $\beta$.

We note that, although the effect of $\beta$—and thus of the regularization—is rather small when considering the averaged normalized MSE alone, we find that the improvement in the structure of the latent-state distribution at the optimal $\beta$ is significant.

Following the approach used for MNIST, we visualize the latent qubit states on the Bloch sphere using the expectation values of the Pauli operators $(P_x, P_y, P_z)$. Since the expectation value of $P_y$ is consistently zero, we restrict the visualization to the $P_x$–$P_z$ plane. For $\beta = 0$ and $\beta = 0.2$, we show the latent distributions corresponding to one selected initialization (out of nine independent runs), specifically the run achieving the lowest MSE for the generative factor position on the aligned qubit.

\begin{figure}[ht]
    \centering
\includegraphics[width=0.9\linewidth]{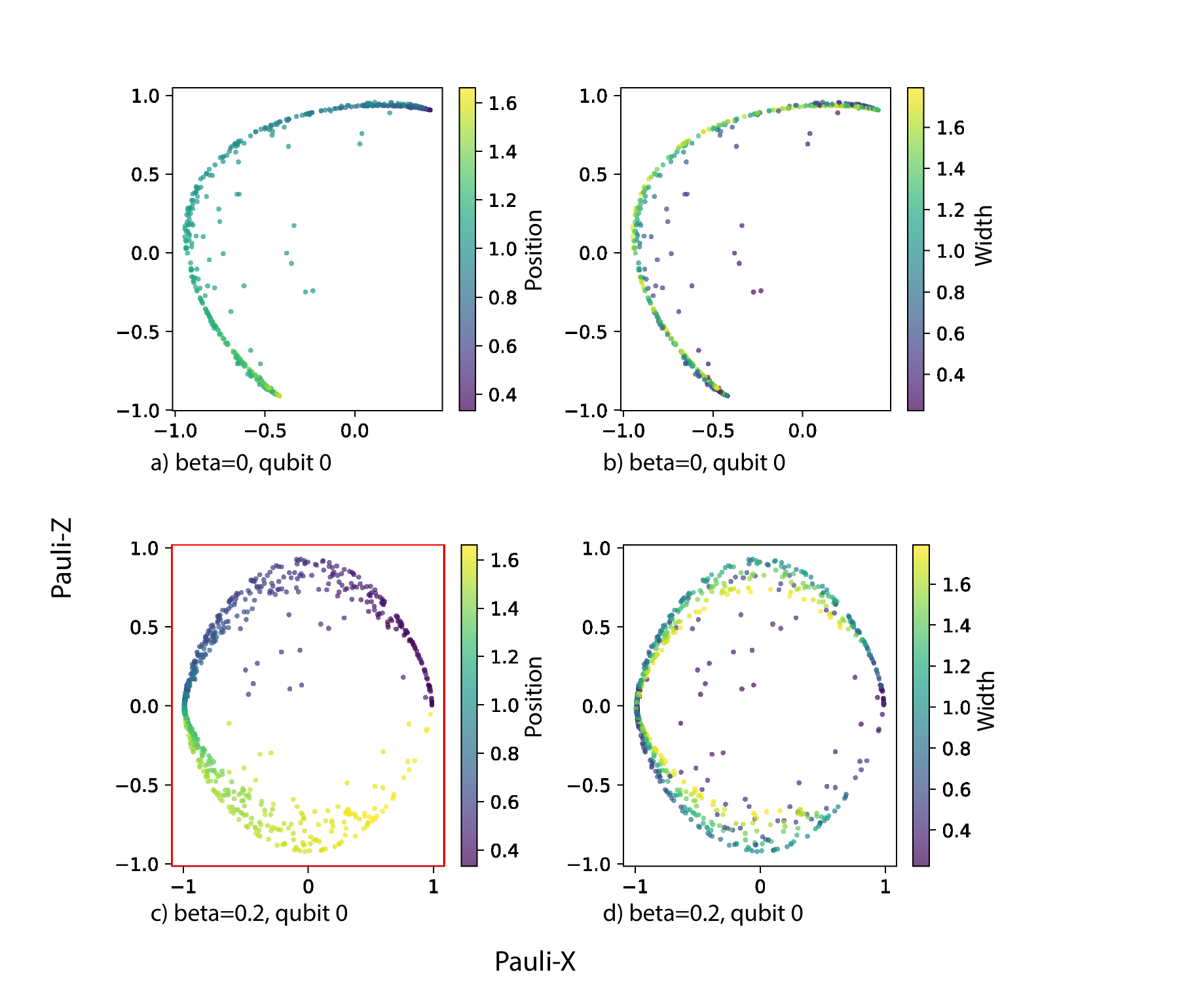}
    \caption{Latent distributions for spectral lines. Qubit 0 aligns with the position factor while showing a much weaker association with the width factor, indicating successful disentanglement of the latent variables. At $\beta=0.2$, the latent distribution becomes significantly more structured compared to $\beta=0$, reflecting the effect of regularization on the learned representation.}
    \label{fig:Spectral_latent_distribution}
\end{figure}

\autoref{fig:Spectral_latent_distribution} panels a) and c) show the distribution of qubit 0 colored by position values (aligned). Whereas panels b) and d) show the distribution of qubit 0 colored by width values (unaligned). At $\beta=0$, the MSE between qubit 0 and position is 0.0123, whereas for the unaligned factor width it is 0.1548. At $\beta=0.2$, these values become 0.0097 for position and 0.1704 for width, respectively. This confirms that qubit 0 encodes substantially more information about position than about width. The figures directly illustrate that the position factor is successfully embedded in this qubit. This also shows that there is little  correlation between qubit 0 and the factor width, consistent with the fact that width information is primarily encoded in qubits 1 and 2.
The latent-state distribution in panel c) at $\beta=0.2$ shows a clear improvement in both latent-space coverage and structural organization. This highlights the beneficial effect of regularization on the learned representation, even though the changes in the averaged MSE appear relatively small. This may indicate that the underlying neural network is indeed too expressive, allowing it to fit the data well even the structural pattern is less clear.

\section{Discussion}
We examined the ability of the quantum variational autoencoder, using a maximally mixed state as the prior, to discover disentangled factors. Our results show that the learned quantum representation, obtained in a fully unsupervised manner, decomposes into independent subsystems, each encoding a meaningful factor in a disentangled form. This represents an important step toward quantum machine learning models that solve tasks by capturing the underlying structure of a problem, analogous to the human ability to decompose complex information into interpretable concepts. These quantum representations reside in Hilbert space and obey quantum probability, making them fundamentally distinct from their classical counterparts.
Future work may focus on investigating quantum embeddings in domains where existing classical models are well established and their behavior is well understood, such as computer vision, protein–protein docking, and natural language processing, to systematically characterize the differences between classical and quantum representation learning.

\printbibliography

\section{Method}

\subsection{QVAE Architecture}

\begin{figure}[ht]
\centering
\includegraphics[width=1\linewidth]{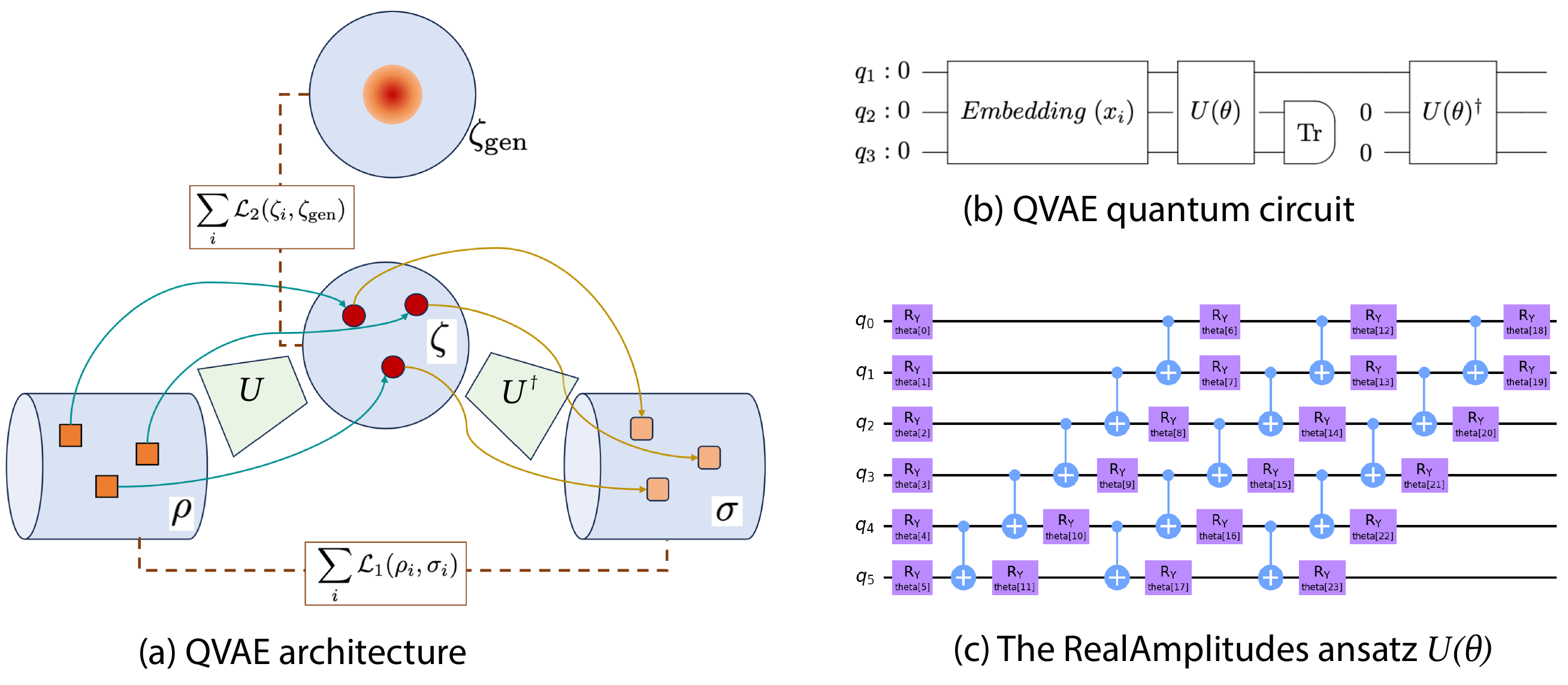}
\caption{QVAE Model architecture.}
\label{fig:encoder_circuit}
\end{figure}

Our model follows the architecture of $\zeta$-QVAE introduced in Ref.\,\cite{PhysRevA.111.042416}, as shown in \autoref{fig:encoder_circuit}\,(a) and (b). It consists of three components: an amplitude-embedding module that maps classical data $x_i$ to quantum states $\rho_i$, an encoder $U(\theta)$ that maps these states into a lower-dimensional latent representation, and a decoder $U(\theta)^{\dagger}$ that reconstructs the input states from the latent representations.
After the encoder’s unitary operation is applied, we trace out the first $N_T$ qubits—referred to as the trash qubits to obtain the lower-dimensional latent state.
For decoding, we restore the dimensionality by taking the tensor product of the latent state with a reference zero state of size $N_T$, and then apply the decoder unitary to produce the reconstructed state.
The decoder employs the same circuit ansatz as the encoder, with parameters fixed to implement the inverse of the encoder transformation. We use the $\verb|qiskit.circuit.library.RealAmplitudes|$ ansatz, which consists of alternating layers of single-qubit $R_Y(\theta)$ rotations and entangling Controlled-X gates. 
\autoref{fig:encoder_circuit}\,(c) illustrates the ansatz for an example circuit with 6 qubits and 3 layers.

The objective function of $\zeta$-QVAE is
\begin{eqnarray}
\mathcal{L} &=& \sum_i \left( \mathcal{L}_1(\rho_i,\sigma_i) + \beta \mathcal{L}_2(\zeta_i,\zeta_{\text{gen}}) \right),
\label{eq_objs}
 \end{eqnarray} 
where the index $i$ runs over all data points. Here, $\rho_i$ denotes the input state, $\sigma_i$ the reconstructed state, $\zeta_i$ the latent state, and $\zeta_{\text{gen}}$ the maximally mixed state used as the prior.
This loss comprises a weighted combination of a reconstruction term $\mathcal{L}_1$ and a regularization term $\mathcal{L}_2$, with the hyperparameter $\beta$ controlling the relative strength of the regularization. Following Ref.\,\cite{PhysRevA.111.042416}, we choose the negative fidelity between the input and reconstructed states as the reconstruction loss, and the Jensen–Shannon divergence between the latent state and the maximally mixed prior as the regularization loss.

\subsection{Dataset}
We use the MNIST dataset of handwritten digits. The original sizes of the images are $28\times 28$ pixels. The first dataset consists of digits 0, 1, 2, 3, and 4, with 200 images per digit. These images are resized to $8\times 8$ pixels using bilinear interpolation.

For the second dataset, we used 500 samples each of the digits 3 and 7. These digits were chosen because they are not rotation-symmetric and thus have a well-defined rotation angle. Since some input digits may be oriented slightly differently, we first correct for these variations to ensure an accurate representation of the rotation angle. Specifically, we align the dataset so that all images of the same class share a common baseline orientation and overall intensity. We use the first digit 3 as a reference and rotate all other 3s to maximize pixel overlap with this reference. We also normalize the intensity values so that the mean pixel intensity matches that of the reference. The same alignment procedure is then applied to the 7s.
Next, we randomly generate a rotation angle between 0° and 180° for each image using a uniform distribution. This rotation is applied to the original $28 \times 28$ pixel images, and the resulting images are then resized to $8 \times 8$ pixels using bilinear interpolation.

In addition to the MNIST dataset, we also generated a synthetic spectral dataset to evaluate our model on data with two known generative factors. Each data point is a 64-dimensional spectral line representing a Gaussian spectrum controlled by two interpretable factors: the peak position ($\mu$), uniformly sampled from $[10, 50]$, and the peak width ($\sigma$), uniformly sampled from $[1, 8]$.
The synthetic spectra are generated according to
$$y_i = \frac{\exp\left[-(x_i - \mu)^2\right]} {2 \times\sigma^2},$$
where $x = \{0, 1, \ldots, 63\}$ represents evenly spaced spectral coordinates. Since our amplitude quantum state encoding scheme normalizes all inputs, we held the overall amplitude scaling constant and did not introduce it as a third generative factor.
The 64-dimensional spectral vectors ($y$) serve as inputs to $\zeta$-QVAE, while the corresponding values of positions and widths provide ground-truth factor labels for model evaluation.

\subsection{Training Details}
In the computational experiments of this work, the following model parameters are used unless stated otherwise. We employ the COBYLA optimizer and set the maximum number of epochs to 60, with a patience of 20 epochs; that is, training is terminated if the training score does not improve for 20 consecutive epochs. The \verb|rhobeg| parameter of the optimizer is set to 1.0. For the RealAmplitudes ansatz, the number of layers, i.e., the number of repetitions of the rotation layer followed by an entanglement layer, is set to three.
\end{document}